\documentclass[lettersize,journal]{IEEEtran}
\usepackage{algorithm, algorithmic, amsbsy, amsmath, amssymb, epsfig, bbm, mathrsfs, multirow, amsthm, comment}
\usepackage{tabu}
 \usepackage[table,xcdraw]{xcolor}
\usepackage{colortbl}
\usepackage{hyperref}
\usepackage{graphicx}
\usepackage{subfigure}
\usepackage{makecell}
\usepackage{caption}
\DeclareMathAlphabet{\mathpzc}{OT1}{pzc}{m}{it}

\definecolor{mygray}{gray}{0.6}
\definecolor{myblue}{rgb}{0.8,0.85,1}
\let\labelindent\relax
\usepackage{enumitem}
\begin{document}

\title{Beyond Wireless Security: Covert Communications in Large Language Model-enabled Edge Networks}
\author{Yuanai~Xie,~\IEEEmembership{Member,~IEEE},~Jiaxin~Chen,~Zhaozhi~Liu

\thanks{Y.~Xie, J.~Chen, and Z.~Liu are with the School of Computer Science, South-Central
Minzu University, Wuhan 430074, China. \emph{(Corresponding author: Zhaozhi Liu.)}}

}

\maketitle

\begin{abstract}
Large language model (LLM)-enabled edge networks (LLMENs) offer mobile users high-quality and low-latency AI-generated content services in the 6G era. However, unlike typical edge networks, LLMENs present unique security challenges due to the inherent complexity of LLMs, their high computational overhead, and continuous interactions with users. Specifically, both frequent user interactions (i.e., queries and responses) over wireless channels and potential electromagnetic information leakage from intensive LLM computations make LLMENs susceptible to various security threats, such as eavesdropping, jamming, prompt poisoning, and prompt injection attacks. Since existing countermeasures against these attacks often incur prohibitive overhead, developing holistic, efficient, and secure privacy protections for LLMENs is crucial. This article first reviews the vulnerabilities of LLMENs, outlines various attacks, and analyzes the drawbacks of existing countermeasures. To overcome these limitations, we propose a covert communications (CC) and computations approach to enhance both the overall security and efficiency of LLMENs. Furthermore, various supplementary solutions are developed to improve the covertness of this framework. Finally, our approach is further evaluated through a case study where the total latency is minimized under stringent communication and computational security requirements. Numerical results demonstrate the proposed approach's effectiveness in enhancing both privacy protection and the execution efficiency of LLM tasks. 
\end{abstract}

\begin{IEEEkeywords}
Covert communications, covert computations, edge network, large language model, wireless security.
\end{IEEEkeywords}


\section{Introduction}
\label{sec:introduction}
Large language model (LLM)-enabled edge networks (LLMENs) \cite{Zlin}, which deploy LLMs on edge servers (ESs) near mobile devices (MDs), have been proposed to reduce latency and bandwidth costs while keeping user privacy as local as possible.
In LLMENs, ESs receive prompts transmitted from various types of MDs through wireless communications, perform computationally intensive LLM inference tasks, and then deliver the inference results back to the MDs. This interaction process, however, is vulnerable to various security threats, such as eavesdropping, jamming, prompt poisoning, and prompt injection attacks \cite{prompt injection, CZhao}. The dynamic nature of LLM interactions, which process input data instantly to generate contextually relevant responses, significantly increases the risk of sophisticated attacks (e.g., prompt injection) targeting this data flow. Furthermore, the computational demands of LLMs, characterized by their deep neural networks, necessitate robust measures to ensure data privacy and integrity. The computational demands also expose the system to unique risks, such as electromagnetic (EM) information leakage \cite{Y-I} from large-scale computations \footnote{Electromagnetic radiation from computational hardware, such as processors or memory modules, can inadvertently leak sensitive information. Attackers can exploit these emissions to intercept confidential data or deduce algorithms, posing significant security risks.}. Such leakage is rare in typical communication systems, but it exacerbates security issues in LLMENs, increasing the risk of various attacks. 

Numerous defense mechanisms have been developed to counter these attacks \cite{CZhao}. Nevertheless, existing countermeasures often fall short in several aspects. Since each countermeasure is typically tailored to a specific type of attack, implementing multiple countermeasures simultaneously is costly and resource-intensive. Furthermore, deploying various countermeasures often necessitates the utilization of shared resources, which can further complicate the privacy and security issues associated with LLMs. For instance, using high transmission power as a countermeasure against jamming attacks might inadvertently facilitate more potent eavesdropping attacks over wireless communications \cite{yxie}. Hence, there is a pressing need to propose a holistic, efficient, and highly secure framework to combat a broad range of attacks on LLMENs. Notably, these attacks on LLMENs occur when the attacker is aware of the LLM data transmission or computation. However, existing research on defense mechanisms in LLMENs does not fully exploit this fundamental insight.

Covert communications (CC) provide an additional dimension of security by concealing the existence of transmissions, distinguishing them from physical layer security (PLS) and encryption technologies that primarily protect content \cite{XChen}. By transmitting at power levels that blend into background noise, this technology not only prevents adversaries from detecting the existence of communication but also serves as a lightweight alternative to traditional security technologies. However, since LLM inference generates significant EM side-channel leakage, even if wireless transmission achieves perfect covertness, an adversary can detect the presence of a covert task by monitoring the ES's computational power profile \cite{Yu2020}, rendering the wireless covertness futile.

Since detecting the existence of wireless transmissions or LLM computations is often the initial step for external attackers planning various attacks on LLMENs, we propose a novel CC and computation framework. However, integrating CC and computations into LLMENs poses significant technical challenges, particularly in managing downlink inference and communication latency, which are crucial for maintaining user experience and system responsiveness. Therefore, this article presents a joint optimization of the transmission power and CPU frequency of the ES to minimize latency in LLMENs. The major contributions of this article are given as follows.
 \begin{itemize}
     \item Unlike typical communication scenarios where CC is solely deployed to protect wireless links, we apply CC to more complex LLMENs. By extending CC beyond the communication domain, our joint CC and computations framework provides a holistic, efficient, and highly secure defense against a broad range of attacks on both wireless transmissions and computational EM leakages.
     \item We propose a dual-domain covertness mechanism that simultaneously conceals wireless transmissions and obfuscates computational power profiles. Specifically, we employ auxiliary techniques such as active obfuscation, which uses rigorous dummy workloads to mimic LLM matrix operations and mask electromagnetic signatures. To the best of our knowledge, this is the first work to propose concrete methodologies for LLM computational covertness.
     \item To address the latency overhead introduced by this covertness mechanism, an overall system latency minimization problem is formulated. We balance security and execution efficiency by jointly determining the transmission power and the ES's CPU operating frequency under covertness and energy constraints.
 \end{itemize}

The article is organized as follows. Section II introduces the advantages and vulnerabilities of LLMENs, as well as the major attack types and their countermeasures. Section III presents the CC and computations framework together with auxiliary covertness techniques. Section IV gives a case study of using our approach for f LLMENs. Finally, Section V concludes the article and outlines future research directions.

\section{Fundamentals of LLMENs}
\subsection{Advantages of LLMENs}
\label{sec:FL_fund}
The emergence of LLMs has spawned numerous groundbreaking applications across various domains, including intelligent dialogue assistants, robotic control systems, and content creation \cite{YChang}. These applications require high-performance and large-scale computational resources, typically provided by cloud computing networks (CCNs) \cite{Zlin}. However, CCNs often struggle with real-time model inference, massive multimodal data transmission, and robust privacy protections due to high latency and centralized data interactions.

\begin{figure}[t!] 
\centering 
\includegraphics[width=7cm]{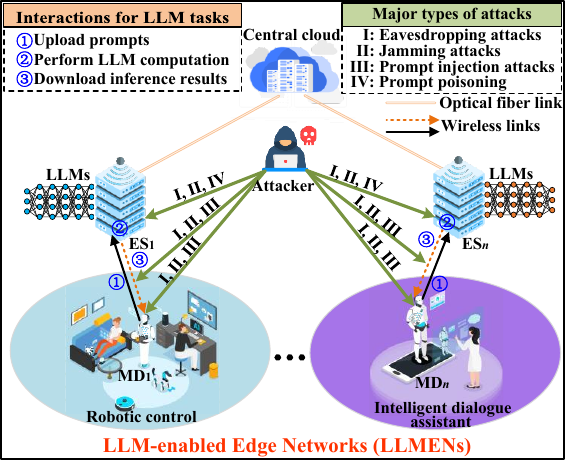} 
\caption{ The architecture of LLMENs and their major types of attacks.} 
\label{fig:1} \vspace{-0.5cm} \end{figure}

To address these challenges, mobile edge computing  deploys LLMs closer to MDs \cite{MXu}. As depicted in Fig. \ref{fig:1}, within a typical LLMEN, MDs wirelessly offload complex user prompts to ESs. Upon receiving these prompts, the ES executes the required LLM inference tasks and subsequently delivers the results back to the MD in multimodal forms. This decentralized paradigm enhances network scalability and reduces communication overhead by distributing resources. It also improves real-time processing capabilities via edge computation and mitigates privacy risks associated with centralized cloud storage.

\begin{table*}[t]
\centering
\scriptsize
\setlength{\tabcolsep}{3pt}
\renewcommand{\arraystretch}{1.15}
\caption{Summary of major security attacks on LLMENs and the corresponding countermeasures}
\label{T1}

\begin{tabular}{|m{0.12\textwidth}<{\centering}|m{0.16\textwidth}<{\centering}|m{0.28\textwidth}<{\centering}|m{0.34\textwidth}<{\centering}|}
\hline
\rowcolor[HTML]{CBCEFB}
\textbf{Attack Types} & \textbf{Sources of Vulnerability} & \textbf{Countermeasures} & \textbf{Drawbacks of Countermeasures} \\\hline
Eavesdropping Attacks & MDs, wireless links, and ES & Encryption methods, PLS & High computational overhead, insufficient security \\\hline
Jamming Attacks & Wireless links and ES & High transmission power \& Active Obfuscation via Dummy Workloads & High energy consumption \\\hline
Prompt Poisoning & MDs and wireless links & TEE, SMPC & \makecell{Limited computational resources and storage space,\\ high computational and communication overhead} \\\hline
Prompt Injection Attacks & MDs and wireless links & RAPT, DPD & Reduced model accuracy \\\hline
\end{tabular}
\end{table*}

\subsection{Vulnerabilities of LLMENs}
Typically, LLMENs are heterogeneous and can be customized for diverse application scenarios, as shown in Fig. \ref{fig:1}. MDs connect to the ES through various wireless links, including terrestrial and satellite networks \cite{BRong}. Similar to other highly mobile environments like vehicular networks, where data exchange via open wireless channels inherently attracts adversaries and various security attacks \cite{ZAlMekhlafi}, this heterogeneity and openness of wireless channels present significant security challenges. Additionally, the unique and complex security issues inherent to LLMENs make them susceptible to various types of attacks. Key vulnerabilities include:

\textit{\textbf{Unsecured Connections:}}
Unsecured connections are a major vulnerability of LLMENs due to their heterogeneity. While advanced encryption and verification techniques can safeguard these connections, they often introduce additional overhead unsuitable for delay-sensitive LLM applications. Consequently, some implementations might reduce the level of security measures for performance, making uploading prompts or downloading inference results vulnerable to interception and manipulation.

\textit{\textbf{Frequent Data Interaction:}} 
LLMs interact dynamically and frequently with input data to generate contextually relevant responses. This frequent interaction increases the risk of sophisticated attacks, such as prompt injection, where malicious inputs compromise the integrity and security of the system.

\textit{\textbf{Exposure of Intricate Computations:}} 
The complex computational tasks involved in LLMs introduce specific vulnerabilities. LLMs' vast number of layers and parameters necessitates substantial computational resources when deployed on ESs. This intense processing not only prolongs inference time but also generates significant electromagnetic emissions. These electromagnetic emissions can inadvertently produce detectable patterns that attackers might exploit. For instance, attackers could analyze these electromagnetic emissions to infer the operations being performed or even the specific nature of the data being processed, posing substantial security risks.

\section{Potential Attacks on LLMENs and Countermeasures}
In LLMENs, various MDs utilize open wireless channels to upload prompts or download inference results from the ES. The openness of wireless channels makes LLMENs vulnerable to various attacks. These attacks primarily
include eavesdropping, jamming, prompt poisoning, and prompt injection. This section provides an overview of these potential attacks and discusses their countermeasures, as summarized in Table \ref{T1}.

\textit{\textbf{Eavesdropping Attacks:}} Eavesdropping attacks pose uniquely severe risks in LLMEN because intercepted wireless prompts and responses often contain semantically rich, highly sensitive contexts (e.g., personal identities and intellectual property). This exposure extends beyond raw data.  Furthermore, the multi-turn continuity of LLM exchanges allows adversaries to reconstruct fragmented interceptions into coherent workflows, elevating the threat above traditional data leakage.

\textit{Countermeasures:}
Encryption and PLS are common defenses against eavesdropping attacks. Encryption methods encrypt transmitted prompts/results using a secret key known by intended recipients, i.e., ESs and MDs. However, encryption suffers from computational overhead and complexity.

While PLS leverages wireless channel randomness and controllable jamming to shield prompts and responses from interception, it cannot guarantee sufficient security, as attackers can still capture key information through computational energy profiling or side-channel analysis of the ES.

\textit{\textbf{Jamming Attacks:}}
Unlike traditional jamming attacks that degrade SINR to cause data loss, jamming attacks on LLMENs induces severe semantic distortion. A major risk is the prompt integrity attack, where interference-induced bit errors alter the semantics of critical keywords in the prompt. For example, a negation like `do not' could be corrupted into `do', triggering unsafe or erroneous LLM responses. The token-by-token generation process is vulnerable to targeted intermittent interference. This gives rise to semantic denial of service attacks, where the adversary truncates or alters generated content during inference to modify meaning without disrupting the entire transmission.

\textit{Countermeasures:}
High transmission power is a commonly adopted countermeasure to mitigate jamming effects. By transmitting prompts/results at higher power, the strength of jamming attacks is relatively reduced, enabling the receiver to capture a higher SINR. Similarly, high CPU frequency is the mainstream countermeasure to prevent jamming from affecting the stability of LLM computations. However, this approach leads to high energy consumption.

\textit{\textbf{Prompt Injection Attacks:}}
Prompt injection attacks manipulate malicious content within prompts or inputs to the model \cite{prompt injection}. The primary aim is to exploit the instruction-following capability of LLMs to induce erroneous behavior or unintended outputs. These attacks can extract or reveal sensitive prompts and confidential information encoded within LLMs. Seemingly innocuous instructions can be strategically formulated to override operational parameters, resulting in disclosure of sensitive prompts or subversion of intended goals. Prompt injection can also be employed for ``polarity poisoning'', a technique that skews the model’s responses by feeding biased prompts, misguiding outputs.

\textit{Countermeasures:}
Privacy-preserving prompt tuning (RAPT) \cite{YLi} and differentially private decoding (DPD) \cite{JMajmudar} are countermeasures to prompt injection attacks.
RAPT enables users to tailor LLMs to private data while preserving privacy. It employs a local privacy setting, applying local differential privacy (DP) techniques to add noise at the individual level, obscuring details during model tuning. RAPT introduces a privatized token reconstruction task alongside downstream tasks, improving task-specific representations. It combines prompt-tuning and prefix-tuning to adjust prompts or add prefixes that guide behavior. Robust and private tuning ensures competitive performance while providing privacy protection.

DPD employs a perturbation mechanism during decoding of a pre-trained LLM, adding noise to protect user privacy. This approach is straightforward and efficient, but compromises accuracy due to the introduced noise. 

\textit{\textbf{Prompt Poisoning:}}
Prompt poisoning refers to inserting malicious content into prompt-related data pathways, including training data and retrieval corpora. Through this approach, an attacker can cause the model to deviate from intended design objectives during inference. These attacks occur during training or data preparation. By contaminating data sources, the model passively absorbs malicious signals and generates harmful, biased, or incorrect outputs.

\textit{Countermeasures:}
During training, resilience against prompt poisoning can be enhanced through multi-layer data security. First, trustworthiness vetting should be reinforced at data ingestion. This is crucial in LLMENs, where data may originate from cloud and edge devices. Measures include verifying traceability of data sources, using data signatures and fingerprints (e.g., hashes and Merkle trees) to ensure integrity, and leveraging data version control and zero-trust policies to govern ingestion of external data \cite{toward}. 

Second, systematic sanitization and anomaly detection must be performed on training data and retrieval corpora to identify contaminated or anomalous content. This includes detecting malicious keywords, samples with statistical deviations, and crafted trigger samples.

Adversarial inference and robustness techniques can be introduced during training to enhance immunity to poisoned samples. Mechanisms include noise injection, adversarial examples, and ``anti-triggers''. Adversarial data augmentation improves robustness against perturbations, while robust training methods mitigate risks of model manipulation, improving reliability in edge environments.

While existing countermeasures, ranging from encryption to data sanitization, provide necessary protection, their high resource overhead limits their viability in resource-constrained LLMENs. Moreover, these defense frameworks are inherently reactive, operating on the premise that an adversary has already detected the communication link.

In contrast, CC serves as a proactive security paradigm by concealing the transmission itself, rendering the signal statistically indistinguishable from ambient noise. This approach fundamentally undermines threat models that depend on signal discovery. Consequently, although CC does not directly filter application-layer attacks, it establishes a critical ``first line of defense'' at the physical layer. By denying adversaries the ability to locate the target link or synchronize with the LLM interaction timing, CC prevents them from initiating targeted injections over the air. By minimizing the probability of detection, it reduces the system’s reliance on computationally expensive reactive measures and functions as an important complementary layer within the cross-layer security architecture.

\section{CC and Computations for LLMENs}
In this section, we detail CC and computations to secure LLMENs. First, an overview of CC and auxiliary covertness techniques are introduced. Then, the superiority and technical challenges of integrating CC into LLMENs are discussed. Finally, a CC and computations approach is developed to ensure the security of LLMENs comprehensively.

\begin{figure}[t!]
 \centering
\includegraphics[width=7cm]{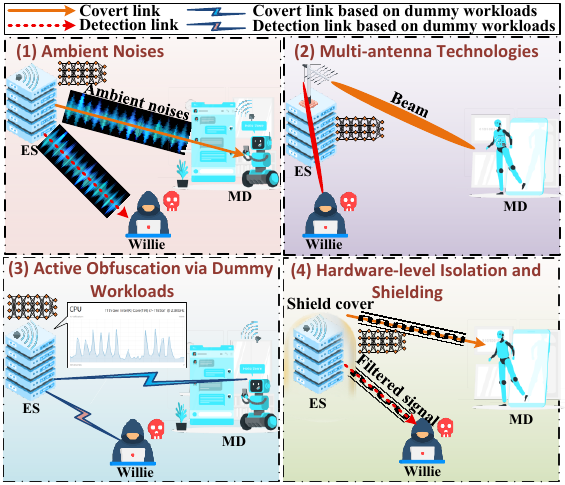}
 \caption{Major auxiliary covertness techniques.}
  \label{fig:2}
  \vspace{-0.5cm}
\end{figure}

\subsection{Covert Communications}
CC is designed to hide the existence of legitimate wireless transmissions from adversaries, guaranteeing a low probability of detection  \cite{XChen}. CC has significant advantages in the following three aspects. Firstly, it provides a cost-effective and low-complexity alternative to hardware-based solutions like TEEs or computationally intensive encryption algorithms. Secondly, unlike PLS, which aims to prevent adversaries from knowing the transmitted messages, CC conceals the very existence of the transmission. This proactive security posture prevents adversaries from launching further attacks and allows CC to counter multiple attacks simultaneously. Thirdly, almost all existing works restrict CC to securing wireless communications (e.g., federated learning networks \cite{yxie}), however, our proposed framework breaks this limitation by extending CC to the computation domain. Recognizing that electromagnetic leakage poses a significant risk of task exposure, CC is integrated to obfuscate the electromagnetic footprint of LLM computations, ensuring that both communication and computation activities remain undetectable.

To address the limitations of prior studies, auxiliary covertness techniques can be integrated into a cohesive framework. These approaches address both communication and computation aspects, ensuring holistic protection for LLMENs. Fig. \ref{fig:2} illustrates these auxiliary covertness techniques aimed at preventing the adversary or eavesdropper, Willie, from detecting the presence of communication and computation. These techniques are given as follows.

\textbf{Auxiliary Communication Covertness Techniques}
\begin{itemize}
    \item \textit{\textbf{Ambient Noise}:} Ambient noise, including background noise and intentional jamming, can be strategically utilized to enhance covertness in LLMENs. When the emitted signal is weak and the cumulative ambient noise is intense, this noise can effectively mask the intended signal. This masking effect enhances covertness by preventing eavesdroppers from detecting the presence of the communication, thus protecting privacy and preventing information leakage.
      \item \textit{\textbf{Multi-antenna Technologies}:}
      By utilizing beamforming and adjusting the signals' amplitude and phase at ESs, multi-antenna technologies can significantly enhance the covertness of LLMENs. Increasing the number of antennas improves beamforming resolution, which allows for more precise targeting of the intended MD and reduces the probability of signal interception by eavesdroppers.   
\end{itemize}

\textbf{Auxiliary Computation Covertness Techniques}
\begin{itemize}
    \item \textit{\textbf{Active Obfuscation via Dummy Workloads}:}
     To ensure the indistinguishability of real and dummy tasks, the dummy workload is not merely a CPU loop but a mathematically rigorous random matrix multiplication.    Since LLM inference relies heavily on matrix-vector multiplication operations, our dummy generator creates random matrices with dimensions matching the target model layers. This ensures that the dummy tasks excite the same hardware logic units and generate spectral power densities similar to legitimate inference, preventing attackers from distinguishing them via signal classification. By maintaining the ES’s CPU at a consistently high utilization state, this method effectively flattens the power consumption profile. By masking the specific electromagnetic signatures of sensitive privacy tasks with continuous dummy workloads, the adversary's SINR is drastically reduced. This ensures that legitimate tasks remain statistically indistinguishable from dummy activities, thereby thwarting side-channel detection attempts.
    \item \textit{\textbf{Hardware-level Isolation and Shielding}:}
    To enhance the covertness of large-scale computations on ESs,  filters and electromagnetic shielding can be deployed to attenuate signal leakages (e.g., acoustic and electromagnetic fields) before they radiate into the environment \cite{Y-I}. Specifically, shielding techniques at the device or room levels (e.g., Faraday cages) can effectively mitigate electromagnetic emissions \cite{RefC}. However, unlike standard video or web traffic, LLMs process vast amounts of sensitive, contextually rich text data with complex interdependencies. This makes them uniquely vulnerable to inference attacks via electromagnetic leakage, as adversaries can reconstruct meaningful information from even fragmented signal leakage. Consequently, stricter electromagnetic shielding and filtering are essential to safeguard the nuanced and sensitive data processed by LLMs, ensuring data integrity and privacy.
    
\end{itemize}

\begin{figure}[t!]
 \centering
\includegraphics[width=7cm]{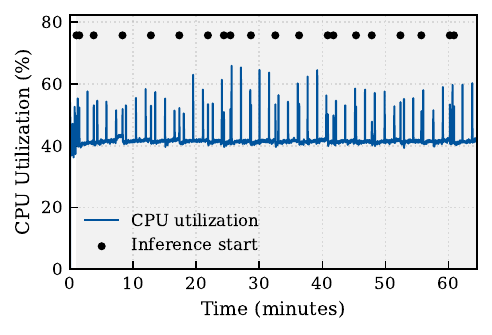}
 \caption{ CPU utilization time series under active obfuscation.}
  \label{cpu_utilization}
  \vspace{-0.5cm}
\end{figure}

To intuitively demonstrate the masking effect of the proposed `active obfuscation' strategy on the computational side, we monitored continuous inference workloads on an ES. As illustrated in Fig. \ref{cpu_utilization}, in contrast to standard operations where inference events trigger distinct, isolated peaks against a low baseline, the injection of auxiliary dummy workloads results in a consistently elevated utilization profile. This generated `artificial noise' effectively masks the characteristic spikes that typically correspond to individual inference instances  (marked by black dots). Consequently, authentic inference events become statistically indistinguishable from background activity, significantly reducing their visibility to coarse-grained external side-channel attacks.

Table \ref{T1} highlights the superiority of our approach over existing countermeasures. However, as shown in Fig. \ref{cpu_utilization}, achieving such high covertness demands sustained resource consumption and often comes at the expense of other performance metrics, such as transmission rate, latency, and available CPU capacity. To address this, the following subsection investigates optimizing computational and communication resources while ensuring holistic protection for LLMENs computations and communications. In particular, we investigate a case study of CC and computations with auxiliary covertness techniques. The transmission power and CPU's operating frequency at the ES are jointly optimized to minimize the overall latency while ensuring a covertness requirement.

\label{sec:FL}
\begin{figure}[t!]
 \centering
\includegraphics[width=7cm]{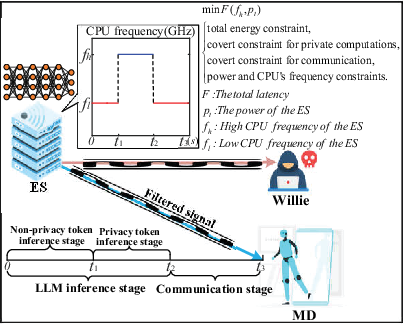}
\footnotesize\caption{Covert communications and computations for LLMENs.}
  \label{F3}
  \vspace{-0.5cm}
\end{figure}

\begin{figure*}[t!] 
\centering 
\subfigure[Total latency versus the ratio $\alpha$ and security threshold $\epsilon$.]{
\includegraphics[width=0.40\linewidth]{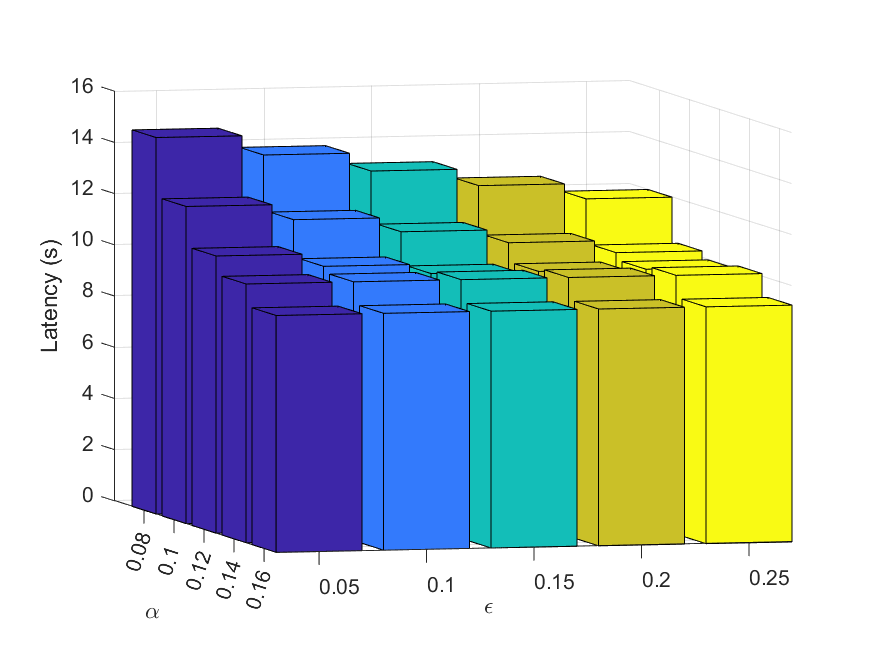}\label{LLMEN:a}} 
\subfigure[Total latency versus security threshold $\epsilon$.]{\includegraphics[width=0.40\linewidth]{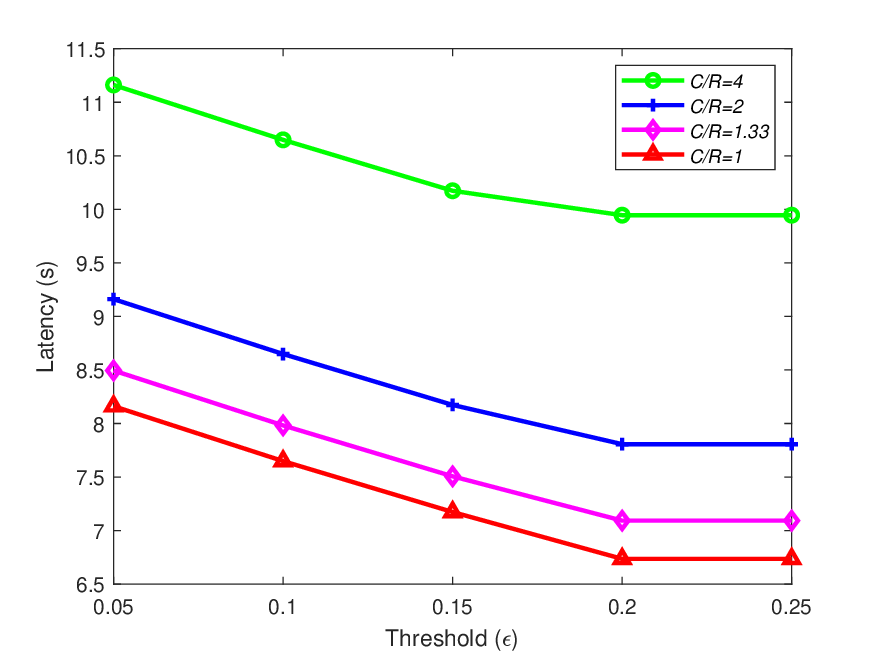}\label{LLMEN:b}} \caption{The MD, ES, and Willie are randomly distributed within a square area of 50 m$\times$50 m. The result data volume $C$ is $20$, the predefined transmission rate $R\in \{5, 10, 15, 20\}$, the maximum computational power and the maximum transmission power of the ES are $21$dBm and $10$dBm, respectively, the preset low CPU frequency $f_l$ has a maximal capacity of $3$ GHz, and the ES has a maximum computational capacity of $5$GHz. The total LLM task volume $T$ is $10^{10}$ bits. The effective switched capacitance of the CPU is set as $10^{-27}$. The conversion coefficient from CPU computation to co-channel electromagnetic interference is 0.02. Note that the latency plateaus when $\epsilon$ increases (e.g., from 0.2 to 0.25), indicating that the system has reached its maximum transmission power and CPU frequency limits, rendering further relaxation of covertness constraints ineffective.} \label{LLMEN} \vspace{-0.5cm} \end{figure*}

\subsection{Balancing Security and Efficiency in LLMENs}
We present a case study of CC and computations for LLMENs. Specifically, we examine the balance between security and efficiency by formulating a latency minimization problem that jointly optimizes transmission power and CPU frequency, considering the covertness and energy constraints of both transmission and inference.

\subsubsection{System Model}
As shown in Fig. \ref{F3}, we consider an LLMEN where an ES processes tasks dominated by privacy tokens and transmits results to a nearby MD, aiming to avoid detection by the attacker Willie. To facilitate the theoretical analysis of latency and covertness, we focus on the autoregressive decoding phase of dense LLMs. In this phase, the computational workload for generating each token is dominated by matrix-vector multiplications, which remain relatively constant per step compared to the prompt processing (prefill) phase.

The ES operates in two distinct modes based on the semantic sensitivity of the input data, categorizing them into non-privacy tokens (e.g., general context or functional words) and privacy tokens (e.g., sensitive user identities or core instructions). Although the intrinsic computational complexity per token is identical in LLM inference, the ES adopts a differentiated resource allocation strategy to balance security and energy efficiency. It uses a lower fixed CPU frequency, denoted as $f_l$, for processing non-privacy tokens to conserve energy, and a higher CPU frequency, denoted as $f_h$, for processing privacy tokens to minimize the time window of electromagnetic exposure. It is worth noting that while the active obfuscation via dummy workloads introduced in Section IV-A provides a macroscopic defense by raising the background noise floor and flattening the CPU utilization profile, sophisticated attackers may still attempt to extract fine-grained token-level information if the sensitive computation persists for a sufficiently long time. By utilizing $f_h$ to compress the execution time window of privacy tokens, the ES reduces the number of electromagnetic samples that an attacker can capture. This microscopic time compression works synergistically with the macroscopic dummy workloads, significantly lowering the attacker’s signal-to-noise ratio and making token-level execution patterns much harder to distinguish from artificial background activity. In this way, dummy workloads conceal the existence of inference at the coarse-grained level, whereas frequency adjustment further suppresses residual token-level leakage at the fine-grained level. When the ES receives a prompt, the server's CPU operates at $f_l$ for non-privacy token inference. Once this stage is over, the CPU works at $f_h$ to perform high-performance privacy token inference. After capturing the result, the CPU switches back to $f_l$ for the communication phase. During this phase, the ES transmits the result to the MD with transmission power $p_t$. Since both LLM computation and the result transmission share a common energy supply on the ES,  both stages are limited by a unified energy budget.

During the inference and communication stages, the ES can decide whether to perform privacy-token-based inference tasks and whether to transmit the obtained results under a predefined event-triggered probability. If the ES decides to perform privacy-token-based LLM inference tasks, but Willie infers that the ES does not perform this task, then a missed detection occurs. If Willie infers that the ES is executing this inference task but the ES does not,  then a false alarm occurs. Specifically, Willie performs detection based on binary hypothesis testing. Willie continuously monitors the electromagnetic emissions and wireless signal energy, comparing the observed statistical features against a decision threshold to determine the presence of legitimate tasks. The detection error probability for this LLMEN is defined as the sum of the false alarm probability and the missed detection probability. A high detection error probability is desired for both stages, ensuring that Willie cannot detect the LLM inference and communication behaviors from the ES effectively. Hence, the detection error probability should be higher than a security requirement $(1-\epsilon)$, where $\epsilon$ is the security threshold. The covert constraints for these two stages are used to achieve holistic protection for the LLMEN.

To minimize the total latency for privacy-token-based inference and result transmission, the ES's transmission power $p_t$ and the CPU's operating frequency $f_h$ should be optimized jointly, while meeting the following constraints:
\begin{itemize}[leftmargin=*]
    \item Total energy constraint for inference and communication stages.
    \item  Covert constraints for the privacy-token-based inference stage.
    \item Covert constraints for the communication stage.
    \item Maximum transmission power of the ES.
    \item  Maximum operating frequency of the CPU.
\end{itemize}
Here, the optimization problem is a standard convex optimization problem. To solve this problem, these two covert constraints are first simplified into deterministic constraints, then dual decomposition is used to solve the simplified problem, and a subgradient-based algorithm is adopted to determine the optimal solution.

\subsubsection{Numerical Results}
We now evaluate the impact of key parameters on total latency. To facilitate the analysis of the fundamental trade-off between latency and security, this case study adopts a simplified single-user scenario; while multi-user scenarios are also feasible, they introduce strong randomness that distracts from the fundamental trade-offs.

Fig.~\ref{LLMEN:a} reveals the impact of the number ratio of non-privacy tokens to privacy tokens $\alpha$ (corresponding to the ratio of their computational consumption) in the prompt and the security threshold $\epsilon$ on the total latency. Under a given $\alpha$, the latency consistently decreases as $\epsilon$ increases. This trend occurs because a higher $\epsilon$ relaxes the restrictions on the maximal CPU frequency and the transmission power, resulting in lower latency. Under a given $\epsilon$, Fig.~\ref{LLMEN:a} indicates that the latency decreases as the ratio $\alpha$ increases when $\epsilon$ is set at 0.05 and 0.1. However, the relationship between latency and $\alpha$ becomes fluctuant when $\epsilon$ varies from 0.15 to 0.25. This fluctuation occurs because, although a higher ratio $\alpha$ (corresponding to a higher CPU frequency $f_h$) can help secure higher-frequency inference based on privacy tokens and higher-power result transmission to some extent, the increase in $f_l$ also generates greater energy consumption and higher background noise. Consequently, this results in increased latencies for both inference and transmission. Hence, there is a key trade-off between the ratio $\alpha$ and the total latency of LLMENs.

Fig.~\ref{LLMEN:b}  shows the impact of the security threshold $\epsilon$ on the total latency under the different ratios of the result data volume ($C$) to the predefined transmission rate ($R$). It is observed that the total latency increases as the ratio of $C$ to $R$ increases. The reason is that a higher $C$ or a lower $R$ consumes more communication time, eventually increasing the latency. However, Fig.~\ref{LLMEN:b} also shows that the total latency remains unchanged when $\epsilon$ increases from $0.2$ to $0.25$. This plateau occurs because the system transitions from a covertness-constrained regime to a resource-limited regime. Specifically, the relaxation of covert constraints allows the ES to increase its transmission power and CPU frequency to minimize latency. However, once the ES operates at its maximum physical limits (i.e., maximum transmission power and CPU frequency), further relaxation of $\epsilon$ cannot yield any additional reduction in latency. Consequently, the total latency stabilizes unless the workload parameters (e.g., the ratio of $C$ to $R$) are altered.

\section{Conclusions and Future Directions}
In this article, we first reviewed existing security issues and corresponding countermeasures during LLM computations and communication stages. To address the limitations of these countermeasures, we proposed a holistic, efficient, and highly secure LLMENs protection framework through CC and computations. We then introduced auxiliary covertness techniques to enhance the security of the computational and communication stages of LLMENs. Furthermore, to balance the security and efficiency of LLMENs, we studied the total latency minimization problem of LLMENs while ensuring that the covert constraints are met. Numerical results were provided to evaluate the performance of CC and computations.

Future research will prioritize addressing the challenges of resource contention in multi-user scenarios. Unlike the single-user case, when a single ES serves multiple MDs, concurrent covert links must share the server's limited power budget and computing resources, leading to a highly coupled and non-convex optimization problem. Additionally, the implications of emerging techniques like mixture of experts and model compression on electromagnetic side-channels warrant deep investigation. While boosting efficiency, their dynamic activation patterns and sparse computations may render traditional constant-power masking ineffective, necessitating content-adaptive covert mechanisms. Finally, distributed covert computing provides a promising research roadmap: splitting inference tasks across multiple devices to diffuse electromagnetic signatures and hinder adversaries from localizing the computation source.

\section{Acknowledgement}
This work was funded by the National Natural Science Foundation of China (Grant No. 62403500) and the Fundamental Research Funds for the Central Universities, South-Central MinZu University (Grant No. CZH25028).


\section*{Biographies}

{\footnotesize 

\noindent\textbf {Yuanai Xie} (IEEE Member) is currently a lecturer at the School of Computer Science, South-Central Minzu University, Wuhan, China. He received the Ph.D. degree in control science and engineering from Yanshan University, Qinhuangdao, China. His current research interests include vehicular networks and covert communications.

\noindent\textbf {Jiaxin Chen} is currently pursuing a Master's degree at the School of Computer Science, South-Central Minzu University. His research interests include LLM inference acceleration and LLM-empowered wireless communication.

\noindent\textbf {Zhaozhi Liu} is currently pursuing a Master's degree at the School of Computer Science, South-Central Minzu University. His research interests include covert communication and LLM-empowered wireless communication.

}
\end{document}